\documentclass[journal]{IEEEtran}
\usepackage{graphicx,
	psfrag,
	epsfig,
	epstopdf,
	amsthm,
	amssymb,
	url,
	subfigure,
	algorithm,
	algorithmic,
	balance,
	enumerate,
	color,
    url,
	setspace,
	tikz,
}
\usepackage[normalem]{ulem}
\usepackage{amsmath}%[centertags][tbtags][sumlimits][nosumlimits][intlimits][namelimits][nonamelimits]
\usepackage[nospace,noadjust]{cite}
\usepackage{pbox}
\usepackage{geometry}
\usepackage{microtype}
\usepackage{multirow}

\newcommand{\cref}[1]{Constraint~\ref{#1}}

\newcommand{\ignore}[1]{}
\newgeometry{left=1.2cm,right=1.2cm,bottom=1.5cm,top=1.5cm}

\begin{document}
%\title{Software Defined Perimeter Framework for Connecting Edge and Cloud Computing} 
\title{Softwarization, Virtualization, \& Machine Learning For Intelligent \& Effective V2X Communications}	
\author{
\IEEEauthorblockN{Abdallah Moubayed, Abdallah Shami}
%\IEEEauthorblockN{Abdallah Moubayed, Abdallah Shami, and Hanan Lutfiyya}

\IEEEauthorblockA{Electrical \& Computer Engineering Department, Western University, London, Ontario, Canada\\ e-mails: \{amoubaye, abdallah.shami\}@uwo.ca		
	%\IEEEauthorblockA{Western University, London, Ontario, Canada \\
	%	e-mails: \{amoubaye, abdallah.shami, hlutfiyy\}@uwo.ca
}\\

%\IEEEauthorblockA{\IEEEauthorrefmark{2} Telus, Toronto, Ontario, Canada \\
%	e-mail: john.bebawy@gmail.com	
%%
}
\maketitle
%\pagenumbering{gobble}
\begin{abstract}
The concept of the fifth generation (5G) mobile network system has emerged in recent years as telecommunication operators and service providers look to upgrade their infrastructure and delivery modes to meet the growing demand. Concepts such as softwarization, virtualization, and machine learning will be key components as innovative and flexible enablers of such networks. In particular, paradigms such as software-defined networks, software-defined perimeter, cloud \& edge computing, and network function virtualization will play a major role in addressing several 5G networks' challenges, especially in terms of flexibility, programmability, scalability, and security. In this work, the role and potential of these paradigms in the context of V2X communication is discussed. To do so, the paper starts off by providing an overview and background of V2X communications. Then, the paper discusses in more details the various challenges facing V2X communications and some of the previous literature work done to tackle them. Furthermore, the paper describes how softwarization, virtualization, and machine learning can be adapted to tackle the challenges of such networks.
\end{abstract}


\section{Introduction}\label{v2x_intro}
\indent The explosion, evolution, and penetration of technology in our daily lives has resulted in increased dependency on connected devices and the emergence of the Internet-of-Things (IoT) concept.  This includes the way we communicate, how we learn, and how we travel from one place to the other. The National Cable \& Telecommunications Association (NCTA) predicted that the number of connected devices will approximately reach 50 billion devices \cite{NCTA_IoT}. Moreover, Cisco projected that the number of mobile-connected devices will reach 11.6 billion by the year 2021 \cite{Cisco1}. This has led to a dramatic growth in mobile data traffic demand which is estimated to reach 49 Exabytes by 2021 \cite{data_growth,Cisco1}.\\ % Furthermore, the GSMA Alliance's report states that the number of unique mobile subscribers will reach 5.9 billion by the year 2025 \cite{gsma_alliance_report}.  This has led to a dramatic growth in mobile data traffic demand which is estimated to reach 49 Exabytes by 2021 \cite{data_growth,Cisco1}.\\
\indent To address this demand, telecommunication operators and service providers have been pushed to upgrade their infrastructure and delivery models. This has led to the emergence of the fifth generation (5G) mobile network system which aims to build on the success of the previous generation (4G) by introducing new network and service capabilities \cite{5GPPP_paper,5G_taleb}. However, 5G is expected to consider various business demands that often have conflicting requirements, which is a divergence from the ``one-fit-all'' model offered by the 4G architecture \cite{5G_taleb}. This will lead to increased innovation and flexibility in terms of the services and programmability of such networks \cite{5G_taleb}. Hence, 5G networks aim to support various use cases that tackle new market segments and business opportunities \cite{5G_usecases}.\\ 
\indent To that end, different paradigms have been proposed. For example, softwarization paradigms such as software-defined networking (SDN) have been proposed as a potential solution for flexible network management. Similarly, software-defined perimeters (SDP) promise to be a core component to secure such networks. Moreover, virtualization paradigms such as network function virtualization (NFV) and cloud/edge computing can help tackle various challenges facing 5G networks including scalability and cost. \\
\indent Furthermore, machine learning (ML) also has a major role in detecting patterns that can help improve the performance and security of modern networks. Particularly, ML can scale well with increasing network size and complexity as the generated data will provide the necessary foundation for the extraction of the updated system characteristics and behavior \cite{ML_networking}. This is further emphasized by the recent studies showing that the use of ML has substantially grew with ML patents filled in the US growing at a compound annual growth rate of 34\% between 2013 to 2017 \cite{ML_statistic1}. \\
\indent One such use case is vehicle-to-everything (V2X) communication. V2X communication has garnered significant interest from various stakeholders as part of the development and deployment efforts of intelligent transportation systems (ITSs) \cite{v2x_interest1,v2x_interest2}. This is due to the many projected benefits it offers including a reduction in traffic-related accidents, introduction of new business models, and a decrease in operational expenditures of vehicular fleets \cite{v2x_access_survey}. V2X communication is required to offer a variety of services ranging from autonomous vehicle operation to traffic flow optimization and in-car infotainment.\\
\indent However, efficient and effective adoption of V2X communication introduces a new set of challenges that is dependent on the service offered and the communication mode adopted. This includes access technology, Quality of Services' (QoS) performance, capacity \& coverage, security \& privacy, scalability \& cost, and standardization. By combining multiple paradigms and technologies, the 5G concept can improve ITS systems by supporting higher network throughputs and lower delays to support basic ITS system services \cite{5G_V2X}. Moreover, 5G networks promise to provide the needed architectures to efficiently manage the different technologies and business models through paradigms such as softwarization, virtualization, and ML.\\
\indent To that end, this work aims to contextualize the challenges facing V2X communication networks and elaborates on the potential methodologies that can address them. Accordingly, this paper:
\begin{itemize}
	\item Discusses in detail the various challenges facing V2X communications and some of the previous literature work conducted to address them.
	\item Describes the role of softwarization, virtualization, and machine learning paradigms and techniques in tackling these challenges and proposes the architecture to implement them.
\end{itemize}

\indent The remainder of this paper is organized as follows: Section \ref{v2x_challenges} provides a brief background about V2X communication modes and applications. It then delves deeper into the different challenges and requirements facing each of the different communication modes. Section \ref{v2x_solutions} presents some potential solutions and methodologies to address the aforementioned challenges. Finally, Section \ref{v2x_conc} concludes the paper.
\section{Current Challenges Facing V2X Communications}\label{v2x_challenges}
\begin{figure}[!h]
	\centering
	\includegraphics[scale=.42]{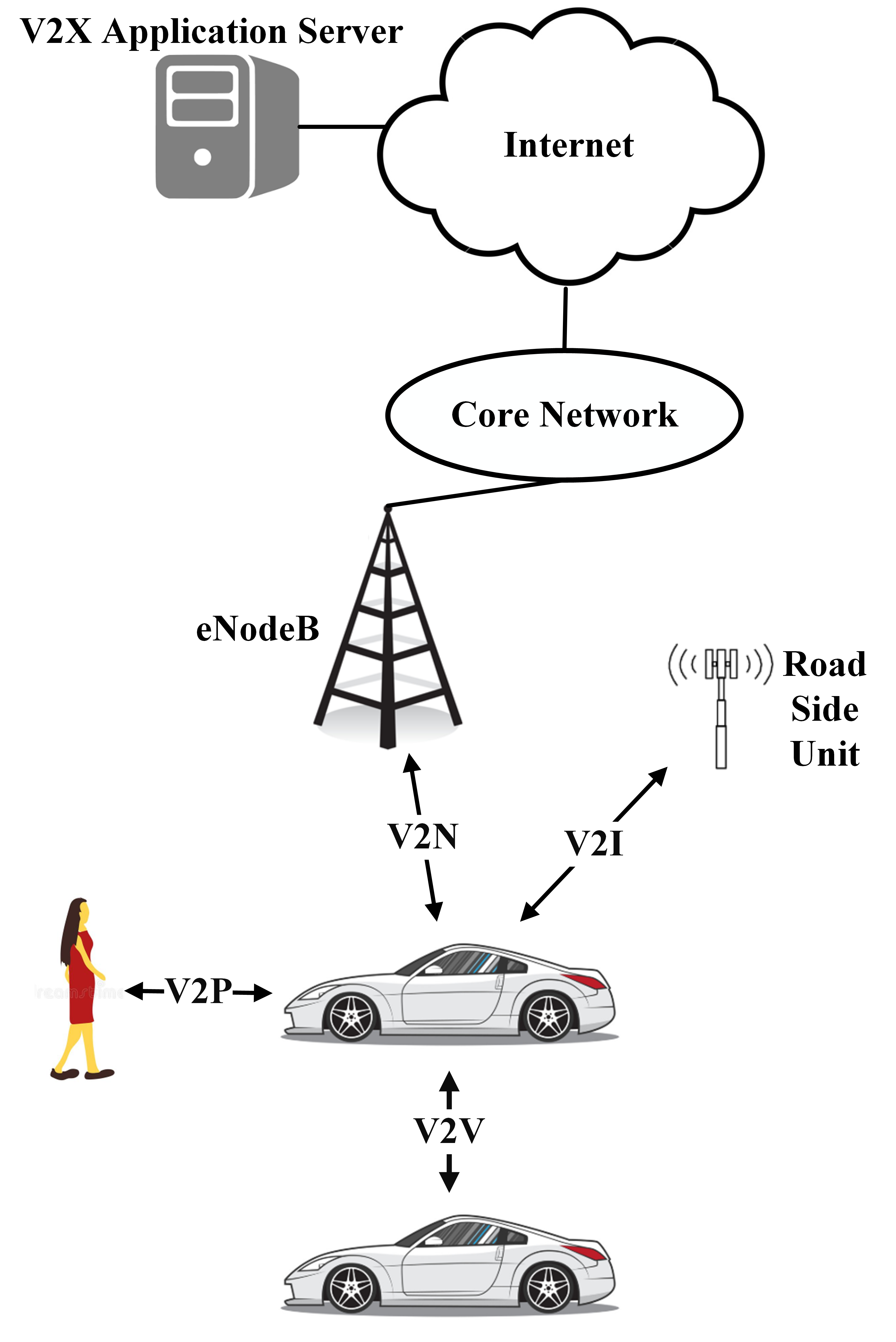}
	\caption{V2X Communication Modes}
	\label{comm_modes}
\end{figure}

\indent As mentioned earlier, V2X communication is one component of an ITS that deals with the communication and coordination between vehicles and their environment. More specifically, it refers to the communication between vehicles and other entities typically found on the road such as other vehicles, pedestrians, and infrastructure. This is done to help provide more economical, efficient, and safe autonomous overland transportation.\\
\indent  Fig. \ref{comm_modes} illustrates the four different V2X communication modes proposed by the 3GPP project, namely: vehicle-to-network (V2N), vehicle-to-infrastructure (V2I), vehicle-to-vehicle (V2V), and vehicle-to-pedestrian (V2P) communication \cite{v2x_communication_modes}.  The communication mode is chosen based on the service being offered.

\begin{figure}[!tb]
	\centering
	\includegraphics[scale=.52]{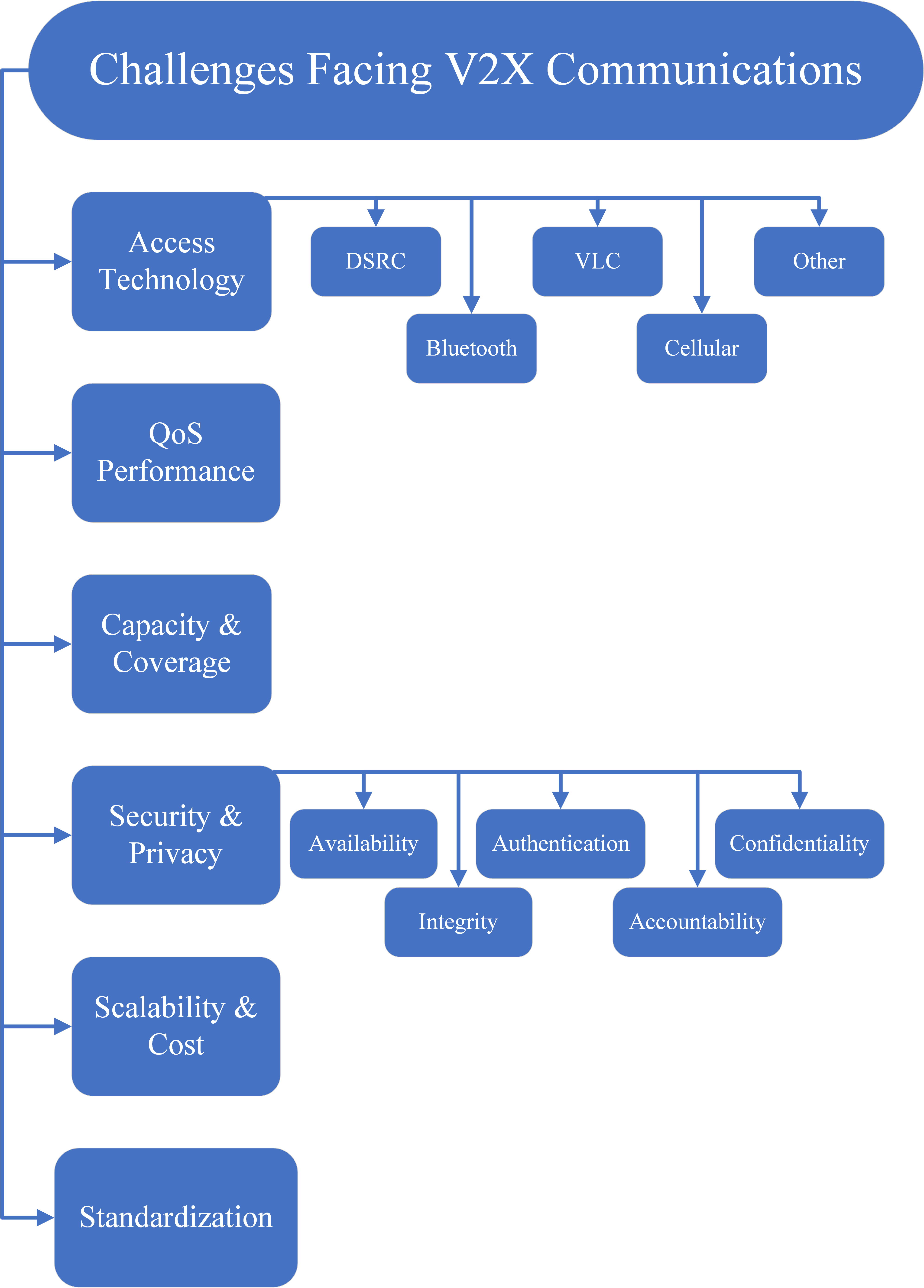}
	\caption{Challenges Facing V2X Communication}
	\label{list_of_challenges}
\end{figure}

\indent Moreover, V2X communications are the basis of a variety of applications and services, each with different requirements in terms of throughput, latency, and frequency. These applications are often grouped into four main categories, namely: autonomous/cooperative driving, traffic safety, traffic efficiency, and infotainment. Table \ref{v2x_background_table} summarizes the different V2X communication modes, the different V2X applications, their description, and their requirements.\\
\begin{table}[!tb]
	\centering
	\caption{V2X Communication Modes and Applications}
	\label{v2x_background_table}
	\scalebox{0.75}{
		\begin{tabular}{|p{3.1cm}|p{4.5cm}|p{2cm}|}
			\hline
			\multicolumn{3}{|c|}{\textbf{V2X Communication Modes}}\\ \hline
			\textbf{Mode} & \textbf{Description}&\textbf{Applications} \\ \hline
			V2N Communication & Communication between a vehicle and a V2X application server (typically using a cellular network) & Infotainment, traffic optimization, traffic safety \cite{v2n_applications1,v2n_applications2}\\ \hline
			V2I Communication&Communication between a vehicle and roadside infrastructure such as road-side units (RSUs)&Traffic safety, information sharing \cite{v2i_applications1,v2i_applications2}\\ \hline
			V2V Communication&Direct communication between two vehicles &Collision warning and avoidance, traffic safety \cite{v2v_applications1,v2v_applications2}\\ \hline
			V2P Communication&Direct communication between vehicles and vulnerable road users (VRUs)& Traffic Safety \cite{v2p_applications1,v2p_applications2}\\ \hline
			\multicolumn{3}{|c|}{\textbf{V2X Applications}}\\ \hline
			\textbf{Application} & \textbf{Description}&\textbf{QoS\newline Requirements} \\ \hline
			Autonomous Driving\newline Cooperative Driving & Focuses on V2V communication, especially between vehicles in close proximity to avoid any accidents &- Throughput $\geq$ 5 Mbps\newline - Latency $\leq$ 10 ms \cite{autonomous_driving1,autonomous_driving2}\\ \hline
			Traffic Safety&Adopts a more general view to reduce the number and severity of inter-vehicle collisions and protect vulnerable users&- Throughput up to 700 Mbps \newline - Latency $\in$ [20-50] ms \cite{traffic_safety1,traffic_safety2,traffic_safety3}\\ \hline
			Traffic Efficiency&Focuses on various tasks such as coordinating intersection timings, planning the route from source to destination for various vehicles, and sharing general information including geographical location and road conditions &- Throughput $\in$ [10-45] Mbps \newline - Latency $\in$ [100-500] ms \cite{traffic_safety2} \\ \hline
			Infotainment&Direct communication between vehicles and vulnerable road users (VRUs)& - Throughput $\approx$ 80 Mbps \newline - Latency $\leq$ 1 sec \cite{autonomous_driving1,traffic_safety2}\\ \hline
	\end{tabular}}
\end{table} 
\indent Despite the variety of applications and services that V2X communication offers, providing an efficient and effective adoption of it introduces a new set of challenges that is dependent on the services offered and the communication mode adopted. A more detailed discussion of these challenges and the previous work done to address them is given below with Fig. \ref{list_of_challenges} summarizing them.
\subsection{Access Technology:}
\indent The first challenge facing V2X communication is choosing the appropriate access technology. This is because different wireless access technologies are capable of offering the basic functions needed for V2X communication. This includes technologies such as the dedicated short range communication (DSRC) which is based on IEEE 802.11p, Bluetooth, visible light communication (VLC), and cellular technologies (LTE and LTE-A) among others. However, each access technology has its own merits and downfalls which can directly impact different metrics and service requirements. In what follows, a brief description of some of the different access technologies proposed in the literature is given along with their merits and downfalls.
\subsubsection{DSRC/IEEE 802.11p}\mbox{}\\
\indent The IEEE 802.11p is a modification of the IEEE 802.11 WiFi standard to support inter-vehicle communication that was proposed in 2010 \cite{802.11p}. Although it did not specify the exact spectrum to be used, leaving that task to regional bodies, most V2X communication occurs around the 5.9 GHz range. When using the IEEE 802.11p technology, vehicles form an ad-hoc network (also commonly referred to as Vehicular Ad-hoc NETwork (VANET) that dynamically changes with different vehicles continuously joining and leaving the network. This often leads to increased network topology instability, which in turn affects other functions such as routing and addressing \cite{802.11p_instability}. Moreover, the significant Line of Sight (LOS) obstructions often found in urban environments negatively impact the performance of DSRC-based solutions, especially when it comes to message transmission \cite{DSRC_LOS_obstruction}. Also, there are concerns about the lack of proper coordination as it also can negatively impact message transmission due to severe congestion of the ad-hoc network \cite{DSRC_no_coordination}.
\subsubsection{Bluetooth}\mbox{}\\
\indent Bluetooth is another short-range wireless communication standard that has been proposed for V2X communication, mainly for intra-vehicle connection. It has been commonly used to directly connect mobile phones or music players to the car's system and enable the utilization of several functionalities such as GPS navigation and music playing. However, the main short coming of Bluetooth for inter-vehicle communication is the limited communication range it supports. As per the standard, Bluetooth devices can only communicate over distances up to 10 meters \cite{bluetooth_standard}. This limits its adoption given that larger ranges are needed for effective V2X communication. Moreover, the slow connection establishment process typically associated with Bluetooth makes it an unlikely access technology for general V2X communication given the highly dynamic nature of VANETs.
\subsubsection{Visible Light Communication (VLC)}\mbox{}\\
\indent A third access technology that has been proposed for V2X communication is visible light communication, also commonly referred to as VLC. As the name suggests, VLC uses visible light frequencies to transmit and receive data \cite{vlc_def}. VLC is an attractive option due to the extremely high data rate (reaching 10 Gbps \cite{vlc_throughput}) and lack of interference it causes to the over-crowded lower-frequency spectrum. Moreover, the energy consumed by light-emitting diodes (LEDs) is much lower than that used for radio-frequency transmission, making it a more attractive potential technology for V2X communication. However, this technology suffers from two major pitfalls. The first is the annoyance caused to drivers due to the constant flickering of light during transmission. The second is the significant impact of the environment on in. In particular, VLC does not operate in non-LOS conditions and is prone to noise caused by other light sources \cite{vlc_def}, which can limit its use in harsh weather conditions.
\subsubsection{Cellular V2X (C-V2X)}\mbox{}\\
\indent Another option proposed is to use cellular network such as LTE or LTE-Advanced (LTE-A) to support V2X communication. This has been typically referred to as C-V2X. The appeal of using the cellular network is the centralized coordination aspect as this facilitates the resource allocation process for users. Moreover, such technologies are able to achieve an end-to-end delay of less than 100 ms \cite{v2x_history10,Moubayed_LTE_D2D_1,Moubayed_LTE_D2D_2,Aqeeli_LTE,Hammad_LTE}, making it suitable for different V2X applications such as infotainment and traffic efficiency. However, current cellular networks still can not offer delay guarantees as stringent as those required by some V2X communication such as traffic safety and cooperative driving. Another concern is the inconsistent coverage that cellular networks offer, especially in rural and mountain areas. Hence, the current state of cellular networks does not allow them to fully support general V2X communications.
\subsubsection{Other 5G access technologies}\mbox{}\\
\indent Other 5G access technologies have been proposed. This includes millimeter Wave (mmWave), satellite radio communication, and non-orthogonal multiple access (NOMA) \cite{mmwave_for_v2x,satellite_for_v2x,noma_for_v2x}. mmWave has the potential to provide gigabit-per-second data rates \cite{mmwave_for_v2x}. Moreover, it can achieve extremely low delays and latencies \cite{mmwave_for_v2x}. Thus, it can be suitable for ITSs as it can fulfill the QoS requirements of a variety of V2X services. Similarly, satellite communication can provide a wide coverage area and consequently help improve information dissemination \cite{satellite_for_v2x}. On the other hand, the adoption of NOMA for ITSs can reduce the access latency, improve spectrum efficiency, alleviate data traffic congestion, and improve packet reception probability \cite{noma_for_v2x}. \\ 
\indent Despite the advantages that such technologies offer such as the high data rate in the case of mmWave and the wider coverage range of satellite communication, they also suffer from major drawbacks. For example, one major drawback that mmWave suffers from is the strong directionality characteristic and LoS connection requirement. This makes is not suitable for highly mobile environments \cite{mmwave_for_v2x}. On the other hand, satellite radio suffers from the end-to-end latency issue as it is estimated to be in the order of hundreds of milliseconds, making it not suitable to all V2X applications \cite{satellite_for_v2x}.
%\vspace{-0.5cm}
\subsection{QoS Performance:}
\indent A second challenge facing V2X communications is achieving and guaranteeing the QoS performance required. As mentioned earlier, some of the different services that need to be supported by V2X communications have extremely stringent QoS requirements, particularly in terms of the throughput and latency. For example, infotainment services have a relatively high throughput requirement but are delay tolerant. On the other hand, services such as autonomous/cooperative driving and traffic safety are more delay stringent but tend to have low throughput requirements. Therefore, meeting these requirements and offering QoS guarantees is a challenging task. Although this is partially dependent on the access technology used, other factors play a role maintaining QoS performance. \\
\indent Significant work has been conducted in the literature to improve the QoS performance of V2X communication systems and networks. Lianghai \textit{et al.} proposed a hybrid access technology scenario in which the data is sent through the LTE-Uu and PC5 interfaces simultaneously \cite{qos_v2x_1}. Their simulation results showed that the proposed scheme significantly reduced the end-to-end latency when compared to the singe access technology scenario \cite{qos_v2x_1}. Similarly, Ben Brahim \textit{et al.} proposed combining the cellular LTE with DSRC technologies for video transmission for connected vehicles \cite{qos_v2x_2}. The experimental results conducted showed improved reliability in terms of packet and frame delivery \cite{qos_v2x_2}. Meng \textit{et al.} also designed an efficient V2X communication system that combines DSRC and LTE together \cite{qos_v2x_3}. The authors investigated the performance of their prototype in terms of packet loss rate, throughput, and end-to-end delay. The experimental results showed that such a combination can support many of the V2X applications and services such as road information sharing and vehicle collision warning \cite{qos_v2x_3}.\\ 
\indent While these works mainly focused on the access technology to improve QoS performance, other works in the literature focused on the routing protocol to achieve better performance. For example, Lugayizi \textit{et al.} compared the performance of two routing protocols, namely Ad-Hoc on-demand Distance Vector (AODV) and Dynamic Source Routing (DSR) protocols, in VANETs \cite{qos_v2x_4}. The protocols were compared using simulations in terms of throughput and end-to-end delay with the DSR protocol showing more stable performance than AODV protocol \cite{qos_v2x_4}. In contrast, Hashem Eiza \textit{et al.} proposed the use of an evolutionary algorithm, namely ant colony optimization, to determine feasible routes under multiple QoS constraints in a VANET scenario \cite{qos_v2x_5}. In particular, the authors showed through simulation that the proposed routing algorithm delivered packets in accordance with the required QoS constraint while adding higher level of security and robustness to the network \cite{qos_v2x_5}. In a similar fashion, El Amine Fekair \textit{et al.} also proposed an artificial bee colony-based scheme, another evolutionary algorithm, to determine the best routes in a VANET depending on the QoS requirements \cite{qos_v2x_6}. Simulation results showed that their proposed routing algorithm improved the performance in terms of the packet delivery rate, end-to-end delay, and the routing overhead \cite{qos_v2x_6}.
%\vspace{-0.5cm}
\subsection{Capacity \& Coverage:} 
\indent A third challenge facing effective V2X communication is the capacity and coverage of the communication. These metrics are not only dependent on the access technology used, but also on the environment in which V2X communication is used and the applications to be supported. As mentioned earlier, using DSRC as the access technology, traffic congestion is easily achieved for such a short communication radius. Moreover, the coverage range is again limited when using such a technology, especially when LOS communication is needed since this is not always possible in a dynamic V2X communication scenario. On the other hand, when the cellular network is used, the congestion is due to the frequent unicast transmissions from the vehicles to the network via the base station. Furthermore, the coverage suffers and may be inconsistent, particularly in rural and mountainous areas.\\
\indent Several works from the literature have explored various ways to model and improve both the capacity and coverage of V2X communication. Starting off with the issue of capacity, Wu \textit{et al.} studied the probability of collision and interference in V2X communication using IEEE 802.11 protocol \cite{capacity_v2x_1}. Their analysis was based on the transmission back off mechanism employed in this protocol. Moreover, the authors proposed that odd hop nodes in the VANET use one channel while the even hop nodes use a second channel. Their simulation results showed that this almost cuts the collision and interference probability by half and significantly reduce the number of colliding or interfering packets \cite{capacity_v2x_1}. Similarly, Chenguang \textit{et al.} analyzed the channel capacity and outage probability of VANETS using three different channel models, namely Rice channel, partial shadow fading channel, and total shadow fading channel \cite{capacity_v2x_2}. Simulation results showed that the outage probability decreases in low signal-to-noise ratio (SNR) conditions. Moreover, the Rice factor had a negative impact on the partial shadow fading channel while having a positive impact on the total shadow fading one \cite{capacity_v2x_2}. Ni \textit{et al.} performed an interference-based capacity analysis using the Car-following model for a 1-dimensional VANET scenario using IEEE 802.11p protocol \cite{capacity_v2x_3}. The authors derived the probability density function (PDF) of different interference conditions including the worst-case interference scenario. Additionally, the authors derived the average network capacity as a function of both the road segment length and transmission power. Their simulation results showed that VANETs should use a lower transmission power when the interference level is acceptable \cite{capacity_v2x_3}. Wang \textit{et al.} on the other hand derived the asymptotic throughput capacity of a VANET and showed that the achievable uplink throughput per vehicle is in the order of $O(1/log\;n)$ where $n$ is the vehicle population \cite{capacity_v2x_4}. Furthermore, the authors proposed a packet-forwarding scheme that exploits the mobility diversity behavior of vehicles to achieve close-to-optimal network throughput \cite{capacity_v2x_4}.\\ 
\indent Similarly, many previous works from the literature investigated and addressed the issue of VANET coverage. Wang \textit{et al.} analyzed the two-hop downlink coverage scenario by deriving the two-hop downlink connectivity probability \cite{coverage_v2x_1}. This probability took into consideration various important factors including road condition, traffic distribution, and the vehicle capabilities. Simulation results showed an overlap between the analytical model and the simulations. Hence, this model can be used to investigate the impact of other design parameters on the VANET connectivity \cite{coverage_v2x_1}. In contrast, Zhang \textit{et al.} studied the coverage probability and area spectral efficiency of an mmWave-based VANET using stochastic geometry \cite{coverage_v2x_2}. The authors derived an analytical model that takes into consideration the blockage effect, channel fading, and interference. Again, simulation results showed a great fit between the analytical model and Monte-Carlo simulations, meaning that the derived analytical model is fairly accurate \cite{coverage_v2x_2}. While these works focused on deriving analytical models to describe the coverage capabilities of VANETs in different scenarios, other works investigated ways to enhance the coverage area for such networks. Salvo \textit{et al.} proposed a simple forwarding scheme that is dependent on on board units (OBUs) to extend the coverage of typical RSUs \cite{coverage_v2x_3}. This forwarding scheme is dependent on efficient inter-vehicle communication and relied on information available at the forwarding OBUs. Simulation results showed that the RSU coverage range was significantly extended (in the order of 20 times greater range). In contrast, Jafer \textit{et al.} proposed a multi-objective genetic algorithm-based method that tries to improve the network coverage by reducing the number of retransmissions and time required for the information to be disseminated \cite{coverage_v2x_4}. Simulation results showed that the proposed genetic algorithm significantly reduced the number of retransmissions (by a factor of $\sim$4 for the urban environment) and network coverage time \cite{coverage_v2x_4}. 
\subsection{Security \& Privacy:}
\indent Another challenge facing V2X communication is that of security and privacy. Due to the different communication modes utilized, they are prone to various types of attacks such as eavesdropping, jamming, wormholes, application attacks, and monitoring attacks \cite{v2x_security_attacks1,v2x_security_attacks2}. These attacks can have a catastrophic implications such as accidents due to spoofed data or even car hijacking. One such example is the hacking of a Jeep car which was remotely killed on the highway with the driver in it as part of a security breach demonstration \cite{v2x_security_attacks3}. The attack was the result of what is called a ``zero-day exploit'' which allowed the hackers to control the car in a wireless manner via the Internet \cite{v2x_security_attacks3}.\\
\indent In particular, attacks in a V2X communication context are divided into five main types, namely availability, authentication, confidentiality, data integrity, and accountability \cite{v2x_security_attacks4}. Attacks on availability refer to attacks that interfere with the transmission and routing of packets. One such attack is jamming attack in which the communication between two entities is disrupted by a malicious transmitter. Attacks on authentication refer to attacks in which the identity of the node is falsified such as spoofing attacks. Such attacks are dangerous because manipulating the data being reported from vehicles can lead to significant road accidents and possibly death of the passengers \cite{v2x_security_attacks4}. Attacks on confidentiality refer to attacks such as eavesdropping in which malicious users passively collect and extract data from vehicular transmissions. This in turn breaches the drivers' privacy as it allows unauthorized users to track their movement and record their data. Attacks on data integrity refer to the transmission of false data. Although it is similar to authentication attacks, the issue here is that the vehicle may be properly authenticated but the data it is transmitting is being altered and falsified. Last but not least, attacks on accountability refer to the loss of event traceability. In this case, authorities need to be able to track malicious users to revoke their access \cite{v2x_security_attacks4}.\\  
\indent Several research works have tried to address this challenge by proposing various schemes and algorithms for more secure V2X communications. Haidar \textit{et al.} proposed the use of a public key infrastructure (PKI) in which digital certificates are exchanged between the vehicles and RSUs \cite{security_v2x_1}. Their testbed implementation results showed achieved more secure communication at the expense of a larger signaling overhead and a higher end-to-end latency due to the encryption and decryption processes involved \cite{security_v2x_1}. Similarly, Villarreal-Vasquez \textit{et al.} investigate the trade-off between safety, security, and performance of V2X systems \cite{security_v2x_2}. Again, the authors proposed the use of PKI as part of the security measures adopted to protect the V2X system. Moreover, the authors proposed that use of an adaptive model that changes the dissemination mechanism based on the sensitivity of the message, the current safety level of the vehicle, and the contextual parameters of the network. Their experimental results showed that using V2V communication as part of the dissemination mechanism significantly reduces the reaction time of vehicles at the expense of added computations \cite{security_v2x_2}. Ulybyshev \textit{et al.} proposed a role-based and attribute-based solution for access control \cite{security_v2x_3}. The proposed solution supports decentralized and distributed data exchange and hence is suitable for the dynamic nature of V2X systems \cite{security_v2x_3}. The experimental results illustrated that the data is indeed protected. However, the cost is the higher computational time needed for encryption and decryption \cite{security_v2x_3}.\\ 
\indent In a similar fashion, many previous literature works have tackled the issue of privacy preservation in a V2X scenario. For example, Wang \textit{et al.} presented a ring signature-based pseudonym generating scheme that enhanced the privacy and anonymity of legitimate vehicles \cite{privacy_v2x_1}.  This scheme allows for anonymous credential authentication while also providing the ability to track suspicious or malicious vehicles \cite{privacy_v2x_1}. Ullah \textit{et al.} also proposed a pseudonym changing strategy that is velocity-based to protect the location privacy of vehicles \cite{privacy_v2x_2}. Their simulation results showed that the proposed scheme is more effective as it achieved a higher anonymity set size and average privacy strength than that of other location-privacy schemes from the literature \cite{privacy_v2x_2}. In contrast, Zhang and Zhu proposed a privacy-preserving intrusion detection system for V2X communication using collaborative machine learning \cite{privacy_v2x_3}. By using the network security laboratory-knowledge discovery and data mining (NSL-KDD) dataset as part of their experiments, the authors showed that there exists a trade-off between security and privacy of the data exchanged which is impacted by the size of the VANET \cite{privacy_v2x_3}. Similarly, Yang \textit{et al} \cite{privacy_v2x_4} proposed a decision tree-based intelligent intrusion detection system.  Their implementation results indicate that the proposed system has the ability to identify various cyber-attacks in autonomous vehicle networks. Furthermore, the proposed ensemble learning and feature selection approaches enable the proposed system to achieve high detection rate and low computational cost simultaneously.
\subsection{Scalability \& Cost:}
\indent Another challenge facing V2X communication is that of scalability and the resulting cost. This is illustrated by the recent report which stated that nearly 125 million cars with embedded connectivity are expected to ship between 2018 and 2022 \cite{scalability_v2x_1}. Moreover, the global market for connected cars is projected to grow by 270\% by that year  with Germany, United Kingdom, and the United States leading the market in terms of total shipments \cite{scalability_v2x_1}. This increase in number of connected vehicles is associated with an increase in the cost of the infrastructure needed to support it. For example, it is estimated that the cost to install a DSRC-compatible system on a vehicle is between 245 and 347 USD \cite{cost_v2x_1} and that of a DSRC-enabled infrastructure deployment around the 17,600 USD per site \cite{cost_v2x_2,cost_v2x_3}. Hence, the cost of deploying a scalable and effective V2X communication system is a major obstacle at the moment.\\ 
\indent To address these issues, many researchers presented different schemes to better improve the scalability of such systems and consequently reduce the costs associated with it. Rajesh \textit{et al.} for example proposed a scalable reactive location-based routing protocol that reduces the routing overhead in VANETs \cite{scalability_v2x_2}. It was shown through simulation that the proposed routing protocol had a higher packet delivery rate, lower end-to-end delay performance, and a lower routing overhead illustrating its efficiency and scalability. On the other hand, Cao \textit{et al.} proposed a scalable and cooperative medium access control protocol  that supports periodic beaconing over the control channel in VANETs \cite{scalability_v2x_3}. The simulation results showed that the proposed protocol had a higher average goodput and a lower average access time \cite{scalability_v2x_3}. In contrast, Kim \textit{et al.} proposed an effective RSU installation strategy that maximized the spatiotemporal coverage under a limited budget \cite{cost_v2x_4}. This problem was formulated as an NP-hard problem that cannot be solved. However, the authors proposed an approximation algorithm with a performance ratio of at least half that of the optimal one \cite{cost_v2x_4}.
\subsection{Standardization:}
\indent Last but not least, standardization is a major challenge facing V2X communication. This is due to various factors. One main issue is the fact that the standards for 5G communication systems, for which V2X communications is just one of the use cases, are still being developed as they are in their infancy. Thus, it becomes much harder to propose standards for this use case when discussions are still underway over many of the characteristics of the 5G technology. Another issue is the fact that different technologies and standardization bodies exist in different countries. This makes it more challenging to have these bodies reach a consensus and to have to technologies converge.\\
\indent Some standardization efforts have been conducted by different bodies that proposed different enhancements on several components and technologies that should be adopted for V2X communication. For example, the 3GPP consortium proposed a new architecture that defines two independent operation modes for a vehicle, namely a direct communication interface called PC5 and a cellular-based interface named LTE-Uu \cite{standardization_v2x_1}. More specifically, the details of the PC5 interface where discussed as part of the Proximity Services (ProSe) discussion in \cite{standardization_v2x_2}. In contrast, the IEEE proposed the 802.11p standard \cite{802.11p} which is based on DSRC as a communication interface for V2X communication. On the other hand, the European Telecommunications Standards Institute (ETSI) expanded on the basic set of applications and service requirements for V2X communication \cite{standardization_v2x_3,standardization_v2x_4}. However, a consensus has not been reached yet as the technologies, requirements, services, and field trials continue to be developed.\\
\indent Table \ref{summary_chall_efforts} summarizes the challenges facing V2X communication systems and networks. Additionally, it lists some of the previous efforts proposed in the literature to address these challenges.
\begin{table}[!htp]
	\caption{Summary of Challenges and Previous Efforts}
	\label{summary_chall_efforts}
	\scalebox{0.75}{
		\begin{tabular}{|p{3.5cm}|p{7.5cm}|}
			\hline	
			\textbf{Challenge} & \textbf{List of Previous Efforts} \\ \hline
			Access Technology &\cite{802.11p}, \cite{DSRC_LOS_obstruction}, \cite{vlc_throughput}, \cite{v2x_history10}, \cite{mmwave_for_v2x},\cite{satellite_for_v2x} \\ \hline
			QoS Performance &\cite{qos_v2x_1}, \cite{qos_v2x_2}, \cite{qos_v2x_3}, \cite{qos_v2x_4}, \cite{qos_v2x_5}, \cite{qos_v2x_6} \\ \hline
			Capacity \& Coverage& \cite{capacity_v2x_1}, \cite{capacity_v2x_2}, \cite{capacity_v2x_3}, \cite{capacity_v2x_4},  \cite{coverage_v2x_1}, \cite{coverage_v2x_2}, \cite{coverage_v2x_3}, \cite{coverage_v2x_4}\\ \hline
			Security \& Privacy & \cite{security_v2x_1}, \cite{security_v2x_2}, \cite{security_v2x_3}, \cite{privacy_v2x_1}, \cite{privacy_v2x_2}, \cite{privacy_v2x_3}, \cite{privacy_v2x_4}\\ \hline
			Scalability \& Cost & \cite{scalability_v2x_1}, \cite{cost_v2x_1}, \cite{cost_v2x_2}, \cite{cost_v2x_3}, \cite{scalability_v2x_2}, \cite{scalability_v2x_3}, \cite{cost_v2x_4}\\ \hline
			Standardization &\cite{standardization_v2x_1}, \cite{standardization_v2x_2}, \cite{standardization_v2x_3}, \cite{standardization_v2x_4} \\ \hline
	\end{tabular}}
\end{table}
\section{Role of Softwarization, Virtualization, \& Machine Learning in V2X Communications}\label{v2x_solutions}
\indent To provide efficient and safe autonomous overland transportation and facilitate V2X communication, several technologies and paradigms can be adopted. This includes softwarization paradigms such as Software-defined networks (SDN) and Software-defined perimeters (SDP), virtualization paradigms such as cloud and edge computing as well as network function virtualization, and machine learning.\\ 
\indent The trend of adopting these paradigms is highlighted by the recent research that estimates that more than 93\% of organizations are using cloud services in some shape or form \cite{cloud_adoption}. Moreover, projections performed by Microsoft predict that the market size for cloud computing will reach 156.4 billion dollars by the year 2020 \cite{cloud_market}. Similarly, the global market for SDN and NFV is expected to reach 54\$ billion by the year 2022 according to various market research reports provided Indian analysis company ``Markets and Markets'' and Irish market research company ``Research and Markets'' \cite{SDN_NFV_market1,SDN_NFV_market2}. In a similar fashion, the International Data Corporation (IDC) projects that the spending on ML technologies will reach 35.7\$ billion in 2019 and 79.2\$ billion by the year 2022 \cite{ML_statistic2}. Hence, these paradigms have the potential to address many of the challenges facing 5G networks and will be major pillars upon which such networks are built on due to the many technical and economic benefits that they can offer.\\
\indent In what follows, a brief description of these technologies and paradigms is provided along with their role in facilitating and enabling V2X communications as shown in Table \ref{summary_methodologies}. 
\subsection{Role of Softwarization}
\indent The concept of softwarization promises to be a key component and enabler of V2X communications. This is due to the many benefits it offers in terms of the network management,   flexibility, scalability, and security.  In particular, two softwarization paradigms, namely software-defined networking (SDN) and software-defined perimeter (SDP), can play a significant role in V2X-based networks. In what follows, these paradigms are discussed in more details especially in terms of integrating them within the V2X communications architecture.
\begin{figure*}[!tb]
	\centering
	\includegraphics[scale=.5]{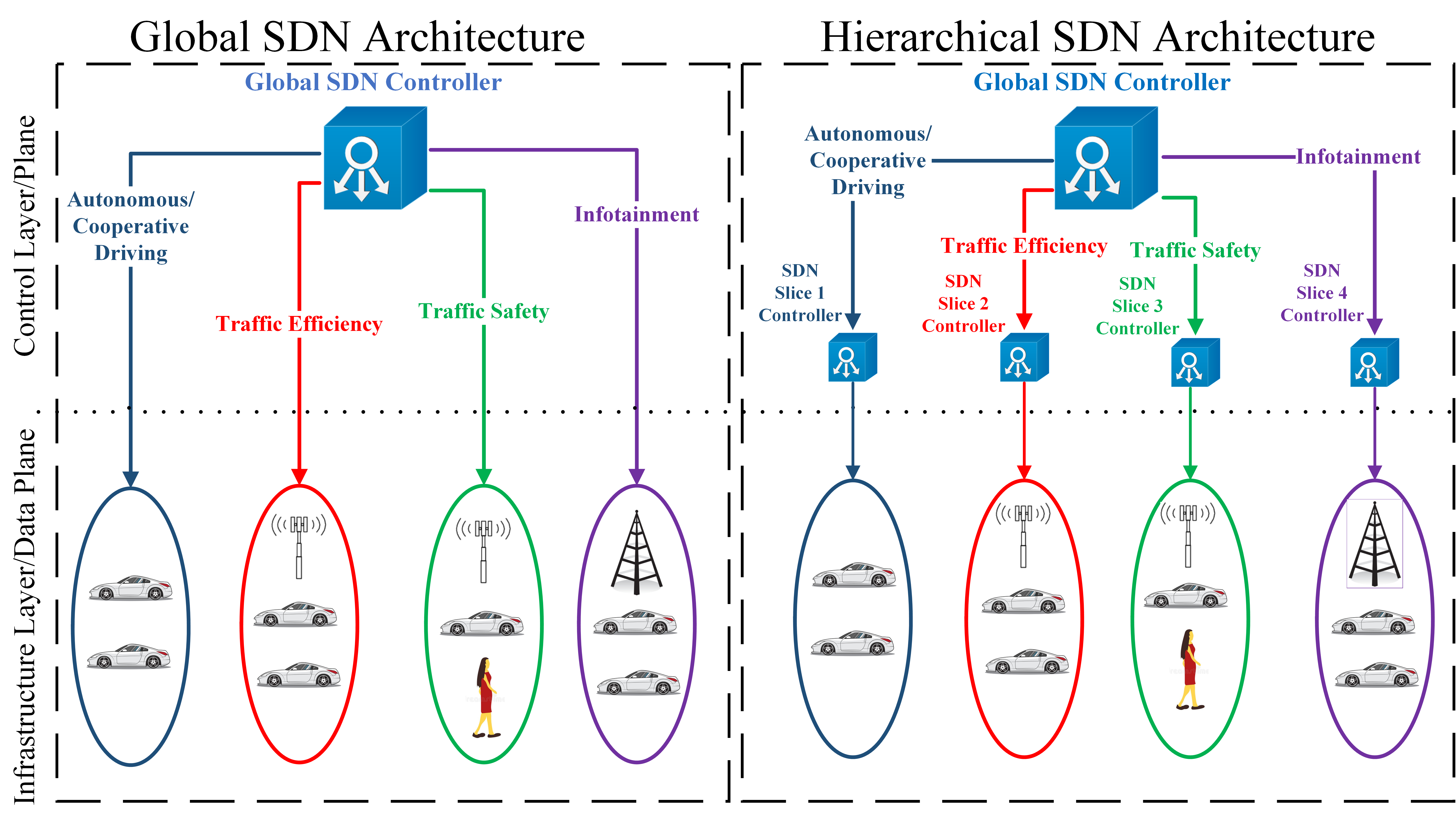}
	\caption{Potential SDN Architectures for V2X Communication}
	\label{sdn_v2x_architectures}
\end{figure*} 

\subsubsection{ Software Defined Networking}\mbox{}\\
\indent \textbf{- Description:} One enabling technology that can play a major role in supporting V2X communications is Software-defined networking (SDN) \cite{sdn_for_v2x,sdn_for_v2x_2}. The concept of SDN decouples the control plane from the data plane using a logical centralized intelligence \cite{SDN1,SDN2}. This facilitates network management and introduces network programmability. Moreover, this offers further flexibility, scalability, and robustness to the network \cite{SDN1,SDN2}.\\
\indent The benefit of adopting such a paradigm for V2X communication is that the different entities such as the vehicles, RSUs, and the infrastructure can all act as SDN switches \cite{sdn_for_v2x_3}. This allows for using a unified interface to manage the vehicular network created and simplifies the integration of heterogeneous networks \cite{sdn_for_v2x_3}. Furthermore, adopting an SDN-based architecture can further support multitenancy in which different government agencies and vehicles' manufacturers can offer multiple services while simultaneously sharing the underlying vehicular network infrastructure in an isolated manner \cite{sdn_for_v2x_4}.\\
\indent Despite the fact that SDN has been proposed in some capacity for V2X communication networks, more can be done to further integrate it as part of a comprehensive solution for an intelligent and effective V2X network implementation. For example, a global SDN controller can be used to manage different network slices corresponding to different applications. As shown in the left hand side of Fig. \ref{sdn_v2x_architectures}, the controller would have a global view of the different network slices. Each slice would represent the set of applications with similar requirements. For example, the system would create four slices, each for one of the application categories mentioned in Section \ref{v2x_challenges}, with the global SDN controller managing both the intra-slice and inter-slice networking. This includes network access monitoring, sharing, and resource allocation decisions. Such an architecture would be more effective in terms of the decisions made due to the complete and global view of all the resources and nodes within the network. \\  
\indent  Another potential architecture is the hierarchical SDN architecture. As can be seen in the right hand side of Fig. \ref{sdn_v2x_architectures}, such an architecture would use a global SDN controller to govern inter-slice management while dedicated SDN controllers can be deployed to manage intra-slice networks. Within such an architecture, the signaling overhead would be distributed between the slice controllers and the global controller.\\ %at the expense of less optimal sharing and resource allocation decisions. \\ 
\indent Such architectures can simultaneously address several challenges. For example, such architectures would facilitate the use of multiple access technologies together, creating a heterogeneous wireless access network. The decision of what access technology to use is made by the SDN controller (global or slice controller) and can depend either on the application requirements, network status (intra-slice and inter-slice), or a combination of both. Also, such an architecture can help address both the QoS performance as well as the capacity and coverage issues by making more informed decisions on network resources sharing and routing paths.\\
\indent \textbf{- Challenge:} One challenge in implementing the SDN concept is the choice of global SDN architecture or the hierarchical SDN architecture. In the case of the global SDN architecture, the amount of signaling overhead generated would be overwhelming, especially given the number of vehicles estimated to be connected. Moreover, such an architecture can have a single point of failure, a characteristic that may be undesirable for service providers. On the other hand, the hierarchical SDN architecture allows for the signaling overhead to be distributed between the slice controllers and the global controller at the expense of less optimal sharing and resource allocation decisions. 
\begin{figure*}[!tb]
	\centering
	\includegraphics[scale=.5]{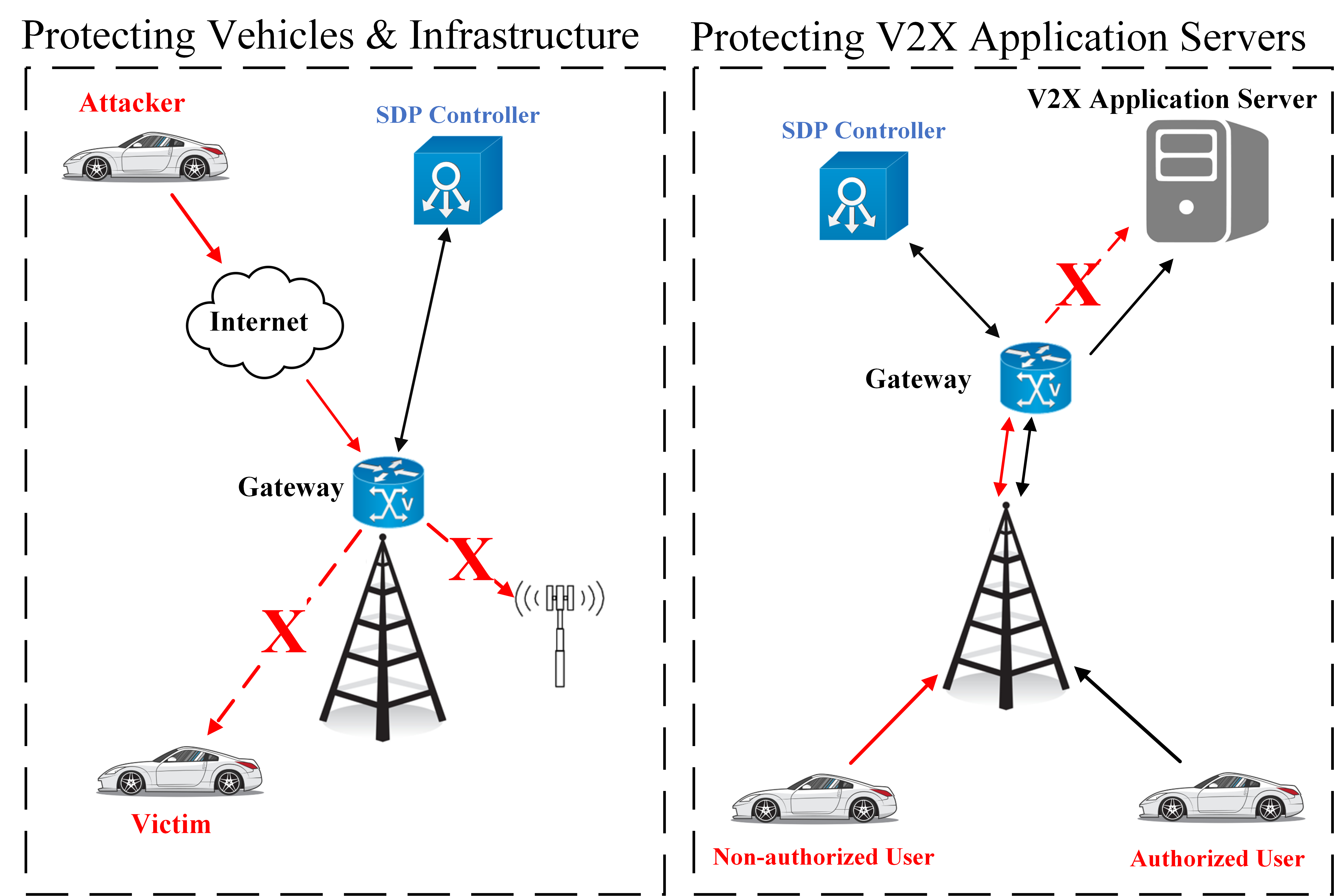}
	\caption{Potential SDP Architecture for V2N \& V2I Communication}
	\label{sdp_v2x_architectures}
\end{figure*}
\subsubsection{Software Defined Perimeter}\mbox{}\\
\indent \textbf{- Description:} Software-defined perimeter (SDP) is a potential solution to tackle many of the security and privacy challenges facing future networks. SDP was originally proposed by the Cloud Security Alliance (CSA) as a security model/framework that can protect networks in a dynamic manner \cite{sdp1,sdp2,sdp3,sdp4}. This concept was developed based on the Global Information Grid (GIG) Black Core network initiative proposed by the Defense Information Systems Agency (DISA) \cite{DISA}. This model follows a need-to-know model where the device's identity is first verified and authenticated before it is granted access the application infrastructure \cite{sdp1,sdp2,sdp3,sdp4}. This makes the infrastructure ``black'' since it can't be detected by users who can only see the infrastructure they are authorized to see \cite{sdp1,sdp2,sdp3,sdp4}. This in turn can help mitigate many network-based attacks such as server scanning, denial of service, password cracking, man-in-the middle attacks, and many others \cite{sdp1,sdp2,sdp3,sdp4}.\\
\indent The SDP concept is built on the notion of providing application/service owners with the power to deploy perimeter functionality as needed to protect their servers. This is done by adopting logical components in place of any physical appliances. These components are controlled by the application/service owners and serve as a protection mechanism. The SDP architecture only provides access to a client's device after it verifies and authenticates its identity. Such an architecture has been adopted by multiple organizations within the Department of Defense in which servers of classified networks are hidden behind an access gateway. The client must first authenticate to this gateway before gaining visibility and access to the server and its applications/services. The aim is to incorporate the logical model adopted in classified networks into the standard workflow. Hence, the SDP architecture leverages the benefits of the need-to-know model while simultaneously eliminating the need for a physical access gateway. The general concept is that client devices are first authenticated and authorized before creating encrypted connections in real-time to the requested servers.\\
\indent In the context of V2X communication, SDP has the potential to offer a more secure environment for all the entities within a V2X network. More specifically, SDP can play a vital role in securing V2I and V2N communications. This is based on the fact that it is very unlikely that a vehicle would attack a neighboring vehicle using V2V communication as this would endanger the attacking vehicle itself. Therefore, V2N and V2I communication offer a more realistic attack scenario for vehicles, similar in fashion to the hijacking attack on a Jeep car discussed earlier \cite{v2x_security_attacks3}. In such a case, the presence of an SDP architecture would have nullified such an attack by placing a gateway at the base station or RSU enabling the V2N or V2I communication mode respectively. By placing the gateway there, as shown in Fig. \ref{sdp_v2x_architectures}, the packets sent by the attackers would be dropped at the gateway before being forwarded to the vehicle, which would deny the attackers access to it and thereby would not allow them to tamper with the car and turn it off for example. Also, any attack by an unauthorized user on infrastructure such as RSUs or traffic signals would also be denied as the default setting of the gateway is to drop all packets from non-authenticated sources. Similarly, any attempt from a non-authorized vehicle to access a service hosted on the V2X application server would be thwarted and stopped at the first hop by being dropped at the access gateway. This is mainly due to the five different layers of security implemented as part of the SDP architecture including the single packet authentication, device validation, and application binding procedures.\\ 
\indent As an example, an architecture similar to that proposed in \cite{sdp3,sdp4} can be adopted (resembling the right hand side of Fig. \ref{sdp_v2x_architectures}). In this case, the entity to be protected is the V2X application server hosting a particular service (for example cooperative awareness basic service). A denial of service attack can be launched targeting this critical service. The SDP gateway placed at the base station or RSU would prevent this attack as it would drop all incoming packets. As a result, based on the results reported in \cite{sdp3} and \cite{sdp4} which can be extended to the V2X scenario, the service can be safely maintained without disruption. Similarly, and by extending the results obtained in \cite{sdp3} and \cite{sdp4}, any port scanning attack aiming at extracting information about adjacent vehicles or infrastructure can also be thwarted  by the gateway. This will result in better security and privacy of the system.\\ 
\indent  \textbf{- Challenge:} One challenge to consider in the case of SDP architecture is its resiliency. In particular, due to the centralized nature of the proposed SDP architecture, it is also prone to single point of failure. Hence, it is crucial that redundancy mechanisms for the SDP controller are applied to ensure that its availability is not compromised.
\subsection{Role of Virtualization}
\indent Virtualization is another concept that promises to be a key enabler and facilitator of effective and efficient V2X communications. This is because it helps improves the flexibility and scalability of the network while simultaneously reducing the associated development cycles and corresponding costs. More specifically, the cloud \& edge computing and the network function virtualization paradigms will offer significant benefits when integrated within V2X networks. In what follows, these two paradigms are elaborated on in terms of how to integrate them into the V2X communications architecture and what are the benefits they can offer.
\subsubsection{Cloud and Edge Computing}
\begin{figure}[!tb]
	\centering
	\includegraphics[scale=.5,trim=0cm 1cm 0cm 1cm]{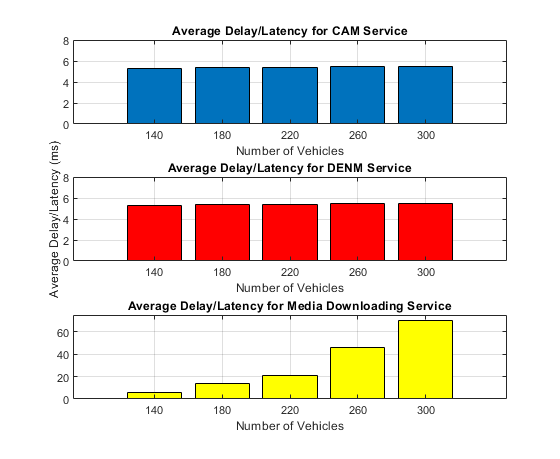}
	\caption{Average Delay/Latency in a Cloud/Edge Computing Environment}
	\label{v2x_mec_example}
\end{figure}

\mbox{}\\
\indent \textbf{- Description:} Two computing paradigms that can play a significant role in V2X communication are cloud computing and edge computing. In layman terms, cloud computing refers to the offering of a pool of virtual and dynamically scalable computing, storage, and memory resources and services to clients on demand over the Internet \cite{cloud_comp1,cloud_comp2}. These services can be offered using various deployment models including Infrastructure as a service (IaaS), Platform as a service (PaaS), Software as a service (SaaS), and Network as a service (NaaS) \cite{cloud_comp1,cloud_comp2}. Due to its characteristics, this paradigm plays a crucial role in the storage and computation of large amounts of data.\\
\indent However, due to the larger potential delays experienced when using the cloud paradigm and the stringent service requirements of technologies such as IoT and V2X communications, the edge computing paradigm emerged as a potential solution \cite{edge_comp1}. Edge computing aims to offer distributed computing and storage at the edge of the network rather than the core \cite{edge_comp1}. This results in lower latencies which is essential to support real-time applications, mobility, and location-aware services \cite{edge_comp1}. \\
\indent Hence, the combination of cloud and edge computing can has the potential to greatly facilitate V2X communication as they can complement each other to support a more comprehensive set of applications and services \cite{edge_comp2}. For example, cloud computing has been proposed to store and respond to queries relating to vehicle records \cite{cloud_for_v2x}. These records can then be used for traffic pattern analysis \cite{cloud_for_v2x}. On the other hand, edge computing was proposed in \cite{edge_for_v2x} in which a V2X service is hosted on the eNodeB rather than in the cloud to facilitate local communication and offer low latency. In particular, this can be suitable for location-related services such as traffic analysis at intersections.\\
\indent Despite some existing efforts that discuss the role of cloud and edge computing in V2X communications, more can be done. One example is offering a coordinated distributed computing platform between the cloud and the edge. In particular, depending on the service required, part of the computations can be done in the cloud for delay-tolerant applications and services while computations for delay-constrained applications should be done at the edge. One example is performing the map update computations in the cloud since the rate of change is not expected to be frequent. On the other hand, computations for speed limits, traffic congestion, and traffic signals can be done at the edge nodes such as base stations and RSUs as these are more delay-constrained and need to be shared with the vehicles in an instantaneous manner. Another example is that of distributed content delivery and caching. This is of particular interest for the infotainment application. In this case, popular videos within a geographical area can be stored locally at edge nodes to reduce the access latency from the users' point of view. Furthermore, information about location-related services such as restaurants or shopping stores nearby can be stored at the edge to offer vehicles some recommendations based on their location.\\ 
\indent \textbf{- Challenge:} One challenge when considering a distributed cloud/edge computing environment is where to place different services/applications. In particular, the placement decision should take into consideration multiple factors including service delay requirement and computing power requirements of the services/applications to be placed. Therefore, the efficient utilization and deployment of cloud and edge computing will have a major role to play in enabling and supporting effective and intelligent V2X communications.\\ 
\indent As an example of the potential of using a combination of cloud and edge computing for V2X services, a preliminary work was proposed in \cite{Moubayed_V2X_placement}. In this work, a distributed cloud/edge computing environment with limited computational resources and supported by LTE-A as the access technology is assumed \cite{Moubayed_V2X_placement}. Using the proposed ``Greedy V2X Service Placement Algorithm'' (G-VSPA), Fig. \ref{v2x_mec_example} shows the average delay/latency of three distinct V2X services, namely Cooperative Awareness Basic Service (CABS), Decentralized Environmental Notification Basic Service (DENBS), and Media Downloading Service (MDS). It can be seen that using such an environment can help improve the performance of these services by reducing the average access delay/latency experienced by users. More specifically, delay stringent services such as the CABS and DENBS had an average access delay/latency of less than 5 ms. This illustrates the benefit of hosting such services at edge nodes to be able to offer lower access delays/latencies and hence better quality of service.
\subsubsection{Network Function Virtualization}
\begin{figure}[!tb]
	\centering
	\includegraphics[scale=.5]{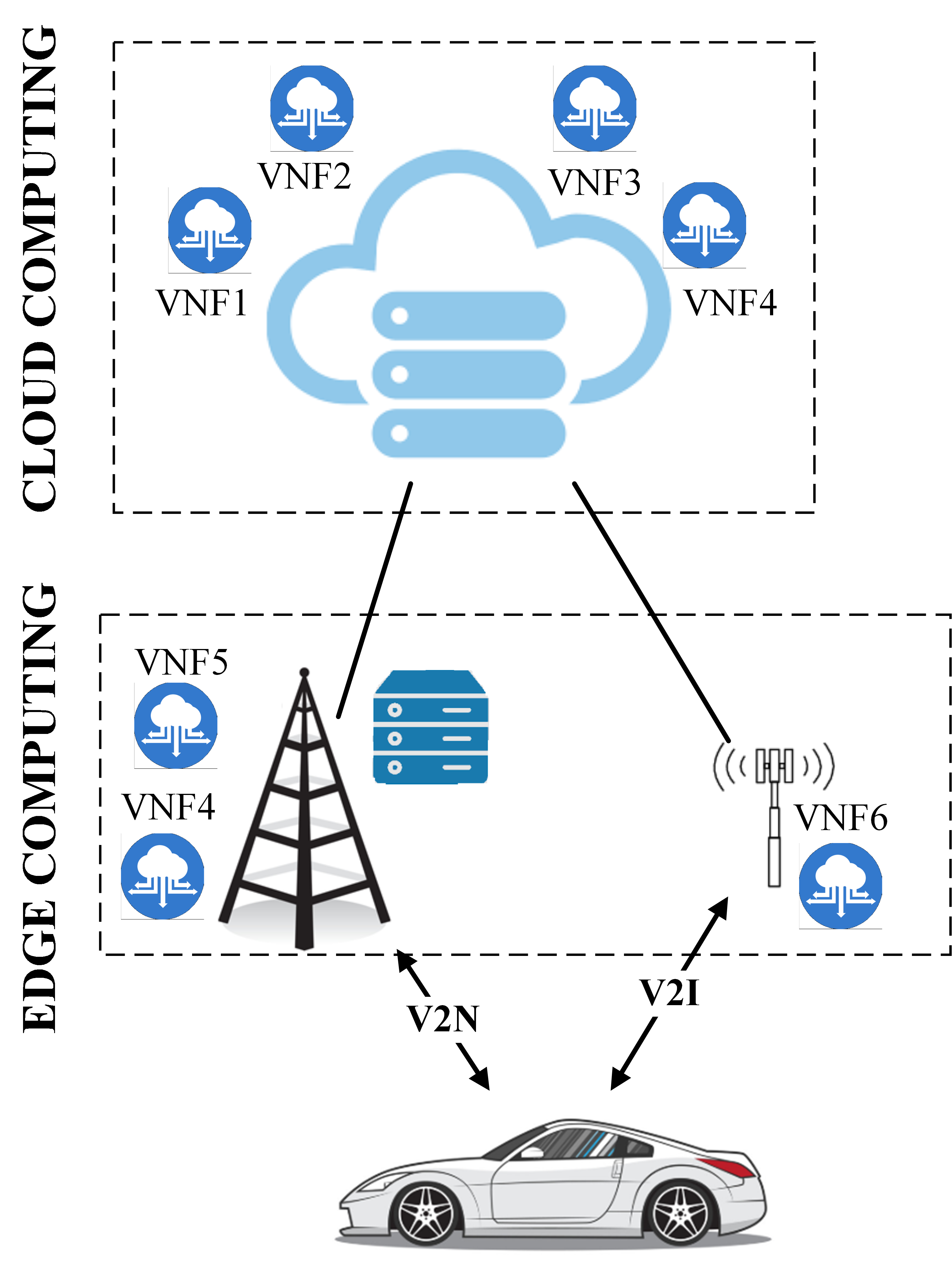}
	\caption{Cloud/Edge Computing \& NFV for V2X Communication}
	\label{nfv_cloud_edge_v2x_architecture}
\end{figure}

\mbox{}\\
\indent \textbf{- Description:} Network function virtualization (NFV) is another paradigm that has huge promise for V2X communications. In essence, NFV refers to the process of migrating network functions from dedicated hardware to software-based applications on standard commercial-of-the-shelf servers \cite{NFV1,NFV2,NFV3}. This is hugely beneficial as this offers more open platforms, better flexibility and scalability, shorter development cycles, and lower capital and operational expenditures (CAPEX and OPEX) \cite{NFV1,NFV2,NFV3}.\\
\indent In the context of V2X communication, NFV can help upgrade and support new services for intelligent on-board systems (IOSs) by virtualizing these IOSs as software applications \cite{NFV_for_V2X}. This helps in making IOSs more efficient as their update becomes easier and faster by using different V2X communication modes \cite{NFV_for_V2X}. Another aspect that NFV can be used to improve V2X communication is by having application-level optimization functions such as load balancing be run on edge nodes to offload some of the burden from the core network. Furthermore, some security functions such as firewalls, intrusion detection systems, and virus scanners can be instantiated and/or migrated from one node to the other on-demand to help improve the overall security of V2X communication. Therefore, NFV has an important role in providing more services for V2X communications.\\ 
\indent For example, extending the architecture proposed in \cite{NFV3} to the V2X environment would be a suitable scenario. In this case, V2X services can be condensed into either micro-services or virtual network functions (VNFs). These micro-services or VNFs can then be dynamically migrated or re-instantiated at different computing nodes both at the core or at the edge. This adds further flexibility and scalability to the system by allowing more effective and efficient orchestration of V2X micro-services and VNFs. Such an architecture is shown in Fig. \ref{nfv_cloud_edge_v2x_architecture}.\\
\indent \textbf{- Challenges:} It is crucial to ensure the high availability of NFVs to maintain the proper functioning of V2X services. This is particularly important for safety-related V2X services and applications given the severe consequences and threat to human life that may be caused by the failure of such NFVs.

\begin{table}[!htp]
	\centering 
	\caption{Summary of Potential Methodologies for V2X Challenges}
	\label{summary_methodologies}
	\scalebox{0.75}{
		\begin{tabular}{|p{2cm}|p{9.2cm}|}
			\hline
			\textbf{Challenge} & \textbf{Potential Methodology}\\ \hline
			\multirow{2}{2cm}{\newline Access Technology} &- SDN can facilitate the use of multiple access technologies by deciding which access technology to use at any moment in time and how to share the available resources depending on application requirements, network status, or a combination of both.\\ 
			&- Supervised machine learning algorithms can be used to determine a suitable access technology based on features such as channel conditions, congestion levels, and performance requirements.\\	\hline
			\multirow{3}{2cm}{\newline QoS Performance} &- SDN can enhance QoS performance by facilitating network resource allocation and sharing as well as making decisions on routing paths.\\
			&- NFV and edge computing can be combined by deploying some of the VNFs in edge nodes for particular applications which would improve the QoS performance.\\
			&- Supervised machine learning algorithms can be used to predict channel conditions or perform rate adaptation. This can help perform more efficient network resource sharing and allocation decisions. \\ \hline
			\multirow{2}{2cm}{\newline Capacity \& Coverage}&- SDN can be used to make resource allocation and management decisions using the Global or Hierarchical SDN controller(s). This can help improve the capacity and coverage by leveraging multiple technologies with the multiple entities such as vehicles, RSUs, and infrastructure acting as SDN switches.\\
			&- Supervised machine learning algorithms can be used to determine a suitable access technology based on features such as channel conditions, congestion levels, and performance requirements. \\ \hline
			\multirow{2}{2cm}{\newline Security \& Privacy}&- SDP architecture can be used for V2N \& V2I communications by placing a gateway at a base station or RSU. This would protect vehicles from attacks through the Internet. It would also protect V2X application servers from unauthorized attackers.\\
			&- Supervised machine learning algorithms can be used to perform intrusion detection or anomaly detection. Using such algorithms, intrusions can be identified and nullified. Furthermore, any vehicle showing anomalous behavior can be identified and investigated for potential threats or malfunctions.\\ \hline
			\multirow{2}{2cm}{\newline Scalability \& Cost}&- Combine NFV and distributed cloud/edge computing to instantiate or migrate VNFs to different locations based on computational requirement, performance requirements, and failures.\\
			&- Supervised machine learning algorithms can be used to predict failure or surges in a specific application or service. This can help in making the instantiation or migration decisions of VNFs from one location to the other. This can help reduce the operational expenditures by reducing costs associated with failure recovery procedures.\\ \hline	
	\end{tabular}}
\end{table}
\subsection{Machine Learning:}	
 \indent Machine learning (ML) and data analytics (DA) algorithms are also major components and enablers of V2X communications. This is due to the fact that such algorithms and techniques can extract information from the available collected data and ``learn'' the behavior without being explicitly programmed \cite{ML1,ML2}. ML algorithms can be decomposed into several categories including supervised learning, unsupervised learning, semi-supervised learning, deep learning, and reinforcement learning \cite{Moubayed_ML,Moubayed_ML2,Moubayed_ML3,Injadat_eLearning}. Firstly, supervised learning refers to the set of algorithms in which a function is learned based on labeled training data \cite{Moubayed_ML}. In this case, training data is composed of a set of
 training instances, each of which is a pair $(x, y)$ where $x$ is an input feature vector and $y$ is the output value \cite{Moubayed_ML}. On the other hand, unsupervised learning refers to the set of algorithms in which a function/pattern is
 learned based on unlabeled training data \cite{Moubayed_ML}. In this case, the training data consists of only inputs $x^1, x^2,...., x^M$ with no known outputs. Hence, such algorithms focus on making sense of the data by finding relations and patterns within it \cite{Moubayed_ML}. Semi-supervised learning combines elements from supervised unsupervised learning algorithms. According, this group of algorithms learn a function/pattern using partially labeled training data \cite{Moubayed_ML}. The goal is to benefit from both the labeled and unlabeled data to get better learning models \cite{Moubayed_ML}. The fourth category of ML algorithms is deep learning. In essence, deep learning can
 be thought of as a large-scale neural network \cite{Moubayed_ML}. In general, deep learning algorithms aim at modeling abstractions found in data using a graph with multiple processing layers \cite{Moubayed_ML}. These processing layers are composed of units (commonly referred to as neurons) that apply linear and non-linear transformations (through activation functions) on the input data to extract useful information from it \cite{Moubayed_ML}. Lastly, reinforcement learning (RL) refers to the set of algorithms in which the action is taken with the goal of  maximizing a cumulative reward metric \cite{Moubayed_ML}. This is typically achieved using a trial-and-error methodology in an attempt to discover the actions with the highest rewards \cite{Moubayed_ML}. The decision taken at a particular time instance often impacts subsequent decisions in addition to the immediate reward achieved. These two features, namely trial-and-error methodology and delayed reward, are the two most distinguishing characteristics of reinforcement learning \cite{Moubayed_ML}. The diversity of ML algorithms make them hugely beneficial in the context of V2X communication since they can help improve and automate several tasks.\\ 
 \indent In particular, machine learning has been used in various ITS-related applications such as vehicle data routing, accident detection, transmit power allocation, and data congestion control \cite{ML_for_V2X_1,ML_for_V2X_2,ML_for_V2X_3,ML_for_V2X_4}. For example, Zhao \textit{et al.} proposed the use of support vector machines (SVM) to generate a routing metric that can improve the data packet delivery rate and the average end-to-end delay \cite{ML_for_V2X_1}. Another example is Dogru \textit{et al.}'s work in which different supervised machine learning algorithms such as artificial neural networks, support vector machines, and random forests are used to detect traffic accidents based on the speeds and coordinates of vehicles \cite{ML_for_V2X_2}. A third example is Gao \textit{et al.}'s work in which a deep neural network was proposed to predict the optimal transmit power that would to maximize the V2X communication system throughput \cite{ML_for_V2X_3}. Lastly, Taherkhani and Pierre proposed the use of unsupervised K-means algorithm to cluster messages as part of a data congestion control unit that determines the appropriate values of transmission range and rate, contention window size, and arbitration interframe spacing for each cluster \cite{ML_for_V2X_4}.\\ 
 \indent In addition to the previous works, there are other works in the literature that adopted ML and DA techniques in the general area of wireless communication and networking that can be adapted to the case of V2X communications. For example, Manias \textit{et al.} proposed the use of decision trees to solve the VNF placement problem in a cloud computing environment \cite{Manias_NFV_orchestration}. The goal was to determine the optimal placement of VNF instances forming a service function chain in such a manner that the delay between dependent VNFs is minimized \cite{Manias_NFV_orchestration}. Another example is Injadat \textit{et al.}'s work in which different ML classification algorithms such as Support Vector Machine (SVM), Random Forest (RF), and k-Nearest Neighbor (k-NN) were optimized and used to identify network intrusion attacks \cite{Injadat_BO}. Similarly, Moubayed \textit{et al.} proposed the use of a majority voting-based ensemble learning classifier to detect suspicious domain names \cite{Moubayed_DNS}. All these proposed models can be adapted to the case of V2X communication systems to address different challenges.\\ 
 \indent Despite some of the previous efforts adopting ML and DA algorithms for ITS-related applications, more can be done. This shows the potential and suitability of of such algorithms in improving V2X communications. For example, supervised ML algorithms such as SVM and RF can be used to determine the suitable access technology based on a variety of features such as channel conditions, congestion level of technologies, and performance requirements of the application. In this case, network service providers would need to use data at the base stations and RSUs about the channel conditions and congestion levels of the different communication technologies to determine the appropriate access technology at that time instance. Another potential use of ML algorithms is to predict channel conditions or perform rate adaptation. To do so, algorithms such as polynomial regression or support vector regression (SVR) can be used to predict the channel conditions based on data collected by the service providers at the base stations and RSUs. This can help make more efficient network resource sharing and allocation decisions. Accordingly, such algorithms can be used to address the ``access technology'' and the ``QoS performance'' challenges. \\
 \indent A second opportunity is using ML models such as decision trees similar to \cite{Manias_NFV_orchestration} to instantiate or migrate VNFs representing different V2X services from one location to the other based on a predicted failure or surge in a specific application or service. Again, all these models can be implemented either at the edge or in the cloud. This allows for a better distributed computation load which in turn reduces the probability of failure. This would address the ``Scalability \& Cost'' Challenge. Another opportunity is applying supervised ML algorithms such as SVM, RF, or k-NN for intrusion detection or anomaly detection similar to those proposed in \cite{Injadat_BO} and \cite{Moubayed_DNS}. Using such algorithms, intrusions can be identified and nullified. In this case, messages sent through the base stations and RSUs would be investigated to identify potential intrusions. Furthermore, any vehicle showing anomalous behavior can be identified and investigated for potential threats or malfunctions. Accordingly, such works would address the ``Security \& Privacy'' challenge. An additional opportunity is determining the optimal speed limit given the perceived road conditions, weather conditions, and traffic level among other factors. Again, regression models can be used to determine these parameters based on historical data collected by the service provider about the traffic levels and road conditions. All these potential research opportunities highlight the potential that ML and DA algorithms have in improving the performance of V2X communication and enhance the performance of ITS systems.\\
 \indent \textbf{- Challenges:} There are multiple challenges to consider when adopting ML and DA algorithms in V2X systems. One such challenge is how frequently to re-train the ML models. This is particularly important given the dynamic nature of such systems which often requires ML models to be regularly re-trained. Another challenge to consider is deciding whether to adopt a centralized or distributed learning architecture. The centralized architecture would provide more accurate predictions since the learning model would have all the data. However, this is associated with higher computational complexity due to the much larger data size. In contrast, a distributed architecture would result in a faster prediction since the model training and learning can be done locally. However, this can come at the expense of a less accurate model. A third challenge to consider is how to optimize the parameters of the ML algorithms. It is known that hyper-parameter optimization can help improve the performance of the ML models. However, the optimization process is associated with added computational complexity. All these factors need to be considered when adopting ML and DA algorithms in V2X communication systems.
\section{Conclusion}\label{v2x_conc}
\indent  The explosion, evolution, and penetration of technology has pushed telecommunication operators and services providers to upgrade their infrastructure and delivery models to meet the increasing demand for data. As such, the 5G mobile network systems have emerged by introducing new network and service capabilities to the previously successful 4G systems. However, 5G is expected to consider various business demands that often have conflicting requirements. Hence, 5G networks are expected to support various use cases that tackle new market segments and business opportunities. To that end, different paradigms have been proposed such as softwarization, virtualization, and machine learning. These paradigms will play a significant role in addressing many of the challenges facing 5G networks, especially in terms of flexibility, programmability, scalability, and security.\\ 
\indent One use case that can benefit from these paradigms is V2X communications.  V2X communication has garnered significant interest in recent years as part of the development and deployment efforts of ITSs. V2X communication is required to offer a variety of services. However, efficient and effective adoption of V2X communication introduced a new set of challenges that is dependent on the service offered and the communication mode adopted.\\ 
\indent To that end, this work put in context the challenges facing V2X communication networks and discussed in detail the potential methodologies that can address them. More specifically, this work described how softwarization paradigms such as SDN and SDP, virtualization technologies such as cloud/edge computing and NFV, and machine learning can be adapted to tackle the aforementioned challenges. Table \ref{summary_chall_efforts} summarized the challenges facing V2X communication systems and networks and listed some of the previous efforts proposed in the literature. Similarly, Table \ref{summary_methodologies} presented the potential methodologies to address them.

\small
\bibliographystyle{IEEEtran}
%\balance
\bibliography{References}
%\vspace{-2cm}
\begin{IEEEbiography}[{\includegraphics[width=1in,height=1.25in,clip,keepaspectratio]{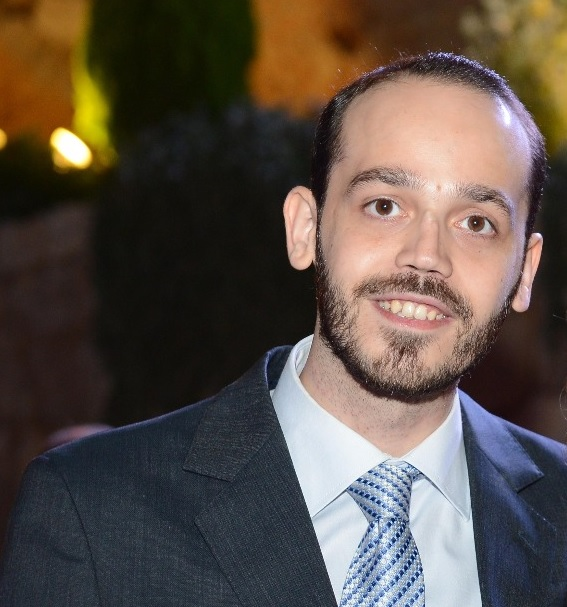}}]{Abdallah Moubayed}
	received his Ph.D. in Electrical \& Computer Engineering from the University of Western Ontario in August 2018, his M.Sc. degree in Electrical Engineering from King Abdullah University of Science and Technology, Thuwal, Saudi Arabia in 2014, and his B.E. degree in Electrical Engineering from the Lebanese American University, Beirut, Lebanon in 2012. Currently, he is a Postdoctoral Associate in the Optimized Computing and Communications (OC2) lab at University of Western Ontario. His research interests include wireless communication, resource allocation, wireless network virtualization, performance \& optimization modeling, machine learning \& data analytics, computer network security, cloud computing, and e-learning. \\
	Postal Address: Amit Chakma Engineering Building, Room 4465, N6A 5B9\\
	email: amoubaye@uwo.ca
\end{IEEEbiography}	
%\vspace{-2cm}
\begin{IEEEbiography}[{\includegraphics[width=1in,height=1.25in,clip,keepaspectratio]{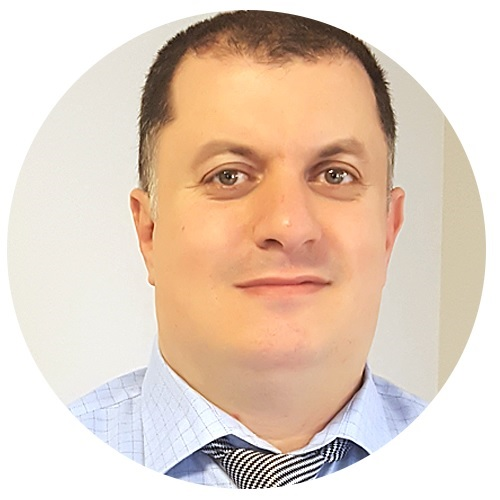}}]{Abdallah Shami}
is a Professor at the ECE department at Western University, Ontario, Canada. Dr. Shami is the Director of the Optimized Computing and Communications Laboratory at Western. He is currently an Associate Editor for IEEE Transactions on Mobile Computing, IEEE Network, and IEEE Communications Tutorials and Survey. Dr. Shami has chaired key symposia for IEEE GLOBECOM, IEEE ICC, IEEE ICNC, and ICCIT. Dr. Shami was the elected Chair of the IEEE Communications Society Technical Committee on Communications Software. \\
Postal Address: Amit Chakma Engineering Building, Room 4455, N6A 5B9\\
email: abdallah.shami@uwo.ca
\end{IEEEbiography}
\end{document}